# Productivity profile of CNPq scholarship researchers in computer science from 2017 to 2021

**MARCELO KEESE ALBERTINI[1] and ANDRÉ RICARDO BACKES[2]**

[1]School of Computer Science, Federal University of Uberlândia, Uberlândia, MG, Brazil

[2]Department of Computing, Federal University of São Carlos, São Carlos, SP, Brazil

## ABSTRACT

Productivity in Research (PQ) is a scholarship granted by CNPq (Brazilian National Council for Scientific and Technological Development). This scholarship aims to recognize a few selected faculty researchers for their scientific production, outstanding technology and innovation in their respective areas of knowledge. In the present study, we evaluated the scientific production of the 185 researchers in the Computer Science area granted with PQ scholarship in the last PQ selection notice. To evaluate the productivity of each professor, we considered papers published in scientific journals and conferences (complete works) in a five years period (from 2017 to 2021). We analyzed the productivity in terms of both quantity and quality. We also evaluated its distribution over the country, universities and research facilities, as well as, the co-authorship network produced.

## INTRODUCTION

Scientific research is central to development of highly specialized human resources in information technology. In Brazil, in all fields, scientific production has increased since mid-2010s mostly fomented by a public policy to expand the universities funded by the federal government. Professors hired during this expansion, due to requirements defined by law, were mostly in tenure tracks, which attracted many new Ph.Ds. willing to pursue science. Due to a greater number of researchers, there have been an increased demand for additional research funds.

A key agency to support researchers is the National Council for Scientific and Technological Development (CNPq), tied to the Ministry of Science, which organizes federal government investments in this field. CNPq has close ties with scientific communities and offers, annually, an important scholarship to those researchers that are most proeminent in each field known as *Bolsa de Produtividade em Pesquisa* (Productivity Scholarship in Scientific Research). This scholarship offers a complementary stipend to the researcher and is regarded as a signal of recognition of excellence in scientific activities.

Currently, productivity scholarships, also known as PQ scholarships, are granted in two levels, related to seniority. In 2022, level 1 was reserved for senior researchers with Ph.D. awarded until 2014 and level 2 to researchers granted Ph.D. until 2019. In recent years, studies have evaluated the profile of grantees of PQ scholarships (Albertini et al. 2019, Fagundes et al. 2020). Our goal was to evaluate the relation of the profile of the grantees in the Computer Science community regarding venue preferences, geographic concentration, distribution by seniority, and bibliometric statistics. We used Qualis index of journals and conference venues and CAPES level of postgraduate programs to cross-evaluate the performance and distribution of productivity scholarships. In addition, we used co-authorship data to observe links among high-productivity researchers.

From our evaluation of distribution of PQ over Brazil, we confirm what has been observed in other fields (Sacco et al. 2016) that there is a very high concentration of PQ researchers in a very small number of states in Brazil (57% of productivity grants is to researchers in only 3 out of 27 states in the Brazilian federation (see Figure 1). When comparing PQ level and productivity, we were surprised to verify that the lower the PQ level, the greater number of publications in journals with higher level of quality (see Figure 2). From co-authorship data, most researchers collaborate little with other grantees. However, we found out a higher collaboration rate between PQ-granted researchers in a few geographically close institutions such as those in São Carlos city (São Paulo state) and in Rio de Janeiro city.

The remainder of this paper is organized as follows: Section 2 describes the current state of the research area. In Section 3 we describe how we collected, selected and structured the data on professors' productivity. Next, in Section 4, we present our analysis over the researchers' productivity. Finally, Section 5 concludes the paper.

## RELATED WORK

Bibliometric assessment is a research area whose main goal is to analyze the bibliographic production to detect and understand the patterns present in it. Its application encloses many topics, such as the analysis of faculty productivity, the analysis of emerging trends and themes, and so on.

The Software Engineering area is the main focus in Wong et al. (2021). The authors systematically collected data from top-quality software engineering venues and compared different years to detect emerging trends and themes. This helped to provide more insights into the software engineering domain. Similarly, the work in Mathew et al. (2017) used topic analysis to detect potential trends in a Software Engineering research community.

In Way et al (2017), the authors analyzed the conventional narrative describing individual faculty productivity trajectories as an obsolescence function. Although the conventional narrative, they found out that the majority of the trajectories are better described by a piecewise linear model composed of 2 linear functions. This study focused on the Computer Science market in North America and a similar study performed using senior researchers' data from the Brazilian Computer Science community confirmed the findings Albertini et al. (2019).

In Bordin et al. (2014), the authors studied the collaboration network from a department. They concluded that many of its metrics, such as the average distance between collaborators, authors who most collaborate, density, and the number of components, could be useful for decision-making at the organizational and individual levels. Similarly, da Silva et al. (2020) also studied co-authorship networks. They focused on academic Brazilian graduate programs in Computer Science and built a network by linking researchers through common publications.

In Fagundes et al. (2020), the authors examine the profile of scientific productivity scholarship researchers (PQ) in the area of Physical Education from 2015 to 2019. They reported a higher concentration of researchers in the southeast region and a higher prevalence for males in the PQ-2 area. Similarly, Sacco et al. (2016) analyzed the profile of 338 PQ scholarships in Psychology from 2012 to 2014. They concluded that only ten universities concentrate 56.7% of researchers awarded with PQ grants, that the southeast region concentrates the highest proportion of PQ grants (55.3%), and are mostly women.

The study in Castioni et al. (2020) analyzed the PQ scholarship distribution in the area of Education. As in other studies, the majority of scholarship holders are in the regions southeast and south, where are most of the federal and public universities in Brazil. In Oliveira et al. (2018), the authors investigate the academic genealogy of the PQ scholarship researchers. The paper is an attempt to map the knowledge propagation through the advisor researcher and the contributions of researchers in the education of human resources.

Similar to our work, Linden et al. (2017) uses bibliometric and social networks analysis metrics to evaluate selected Brazilian Computer Science Graduate Programs. They used Principal Component Analysis (PCA) to compare Programs of levels 6 and 7 at CAPES (a foundation linked to the Brazilian Ministry of Education) and those best ranked in three different international rankings. They concluded that CAPES ranking is different from the worldwide-accepted ones, indicating some kind of "Brazilian Science".

## DATA PREPARATION

Here we detail the data acquisition process we used for this work. We analyzed the productivity of faculty members granted with the Productivity in Research Scholarship (PQ). Granted by CNPq (Brazilian National Council for Scientific and Technological Development), PQ scholarship aims to value researchers having scientific production, outstanding technology and innovation in their respective areas of knowledge. CNPq groups the granted researchers into five levels according to their productivity: PQ-2 (lower classification), PQ-1D, PQ-1C, PQ-1B, and PQ-1A (higher classification).

We consider the last PQ selection notice, i.e., researchers with PQ scholarship starting date on March 1st, 2022. We accessed the researchers' data on March 22nd, 2022. We considered only researchers with no pending issues on that date, thus resulting in a total of 185 researchers in the Computer Science area.

Since we are interested in the scientific production, we collected the Lattes Résumés of the selected researchers, which includes all sorts of academic information (scientific research, master's and Ph.D. students etc.). From each Lattes Résumé we extracted the contribution to the training of human resources (master's and Ph.D. students, undergrad research and term paper) and the list of papers published in scientific journals and conferences (complete works) in a five years period (from 2017 to 2021). For each paper, we collected authors' list, title, journal/conference, year of publication and Qualis index, a Brazilian official classification of journals and conferences maintained by the Coordenadoria de Aperfeiçoamento de Pessoal de Nível Superior (CAPES), a government agency linked to the Brazilian Ministry of Education. Qualis index groups the journals and conferences into nine levels according their relevance: S (no classification), B4, B3, B2, B1, A4, A3, A2 and A1 (higher classification).

## RESULTS AND DISCUSSION

We extracted and analyzed data from 185 researchers/professors from 2017 to 2021, 5 years. Figure 1 shows the number of PQ scholarships granted to these professors. According to CNPq, the higher the classification level of the scholarship, the more is expected of the researcher, not only in the production of quality research but also in the formation of human resources, a research trajectory with impactful results, and peer recognition. Thus, it takes time for a researcher to achieve the productivity and status required in higher levels of the scholarship, as we can see in Figure 2. About 70% of the PQ scholarships belong to the lower classification level (PQ-2) while only 7 PQ-1A (higher classification) were granted (3.78%).

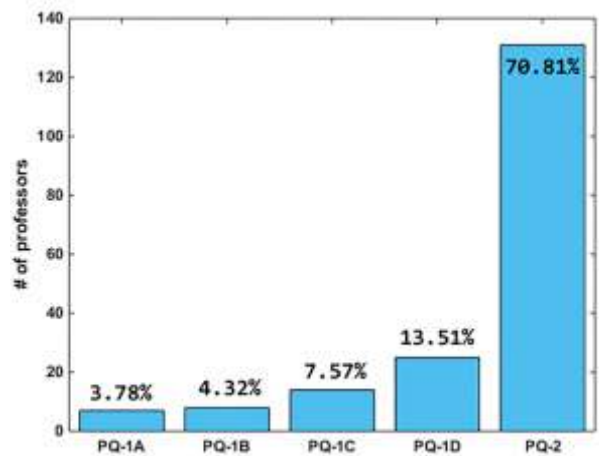

**Figure 1** - Distribution of PQ scholarships.

Selected professors are distributed along a total of 51 universities and research facilities. However, we notice an imbalance in the distribution of scholarships along the universities. According to Figure 3, 24 institutions group 82% (151) of the scholarships granted. In addition, around 90% of these institutions are public, i.e., maintained by federal or state governments. Figure 5 shows the position in a map of each PQ researcher granted. The concentration of researchers in a few regions and/or research centers is impressive, but an unexpected result. The Southeast region of Brazil is characterized by a higher level of development and IDH. As a result, it concentrates 107 (58%) of the scholarships granted and the majority of scholarships in higher levels (Figure 4). Nevertheless, even inside this region, we notice an imbalance, since one of its states (Espírito Santo - ES) has a single scholarship. North region is the most underrepresented and has only one of its seven states represented (Amazonas - AM) with a single PQ scholarship. The Northeast region appears well represented in this last PQ selection process with a total of 41 (22%) grants, including some in higher levels, with special attention to Pernambuco state (PE). States in this region usually present a lower IDH in comparison to Southeast and South regions, so a larger number of grants may be a result of additional investments in education and research over the years.

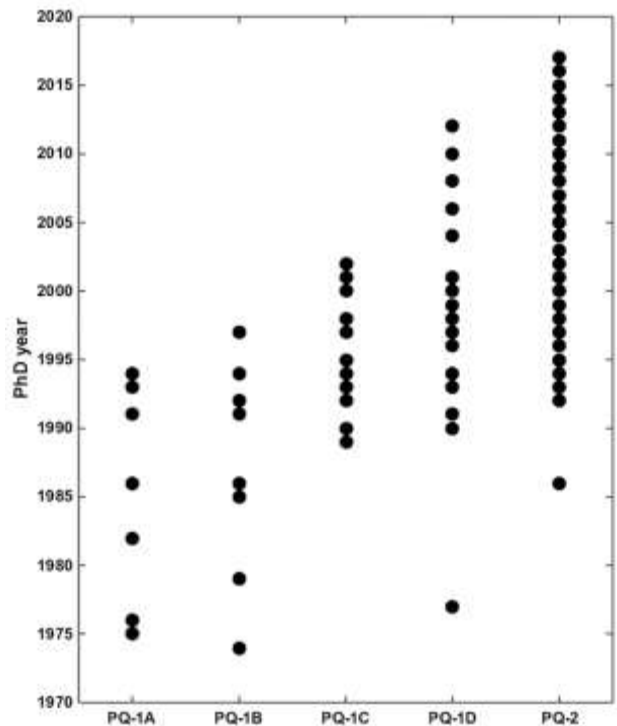

**Figure 2** – Relation between year of obtaining the Ph.D. and the PQ scholarship level.

Every four years, CAPES (Coordenação de Aperfeiçoamento de Pessoal de Nível Superior) carries out a Quadrennial Assessment of postgraduate programs at Brazilian universities. This evaluation considers several factors, such as the scientific production of the supervisors, the quality of the scientific production of the master's and Ph.D. students, and the social impact of the program, which can be technological, economic, and educational or even consist of reducing the indicators of social inequality or in improving the Human Development Index - IDH. CAPES attributes the postgraduate programs the following levels: S - no classification; 1 - weak; 2 - deficient; 3 - regular; 4 - good; 5 - very good; 6 and 7 - world-class excellence.

Figure 6 shows that most of the researchers granted a PQ scholarship belong to a level 7 program (39%). When we consider only the level 1 PQ scholarships (PQ-1A, PQ-1B, PQ-1C, and PQ-1D), the proportion of scholarships belonging to a level 7 CAPES rises to 63%. It is an expected result that a highly evaluated program concentrates a larger number of granted scholarships. However, we also expected that the total number of scholarships followed a linear tendency as we decreased the CAPES level. Contrarily, programs at levels 3, 4, and 5 presented more researchers with granted scholarships than level 6. This may be an indication of the effort of the researchers in these programs to elevate the level of the postgraduate program in the next Quadrennial Assessment.

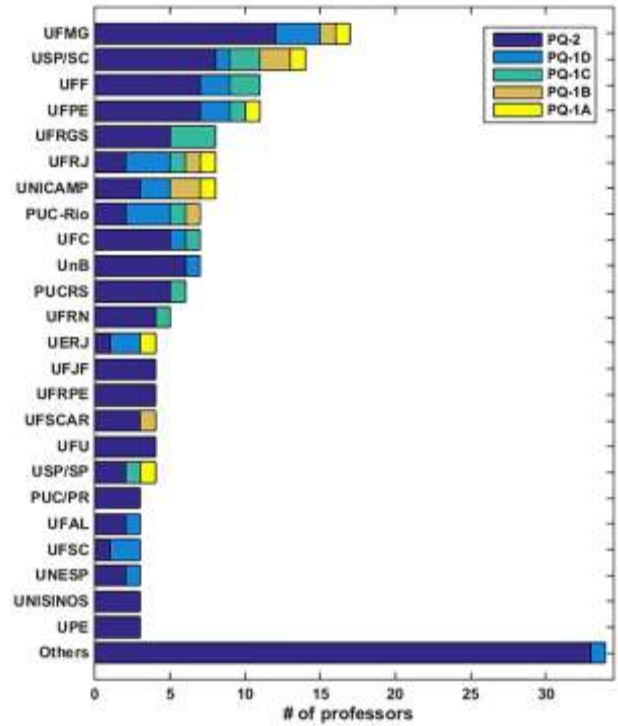

**Figure 3** – Distribution of PQ scholarships granted by universities and research facilities.

Figure 7 shows the average and median publications according to the Qualis index and PQ scholarship. We notice a difference in the Qualis index when we split papers into journals and conferences. In general, researchers tend to target journals with a higher Qualis index to publish a paper. For conferences, the distribution is more homogeneous, lacking any strong indication of selectivity regarding the quality of the conference. An interesting point to notice is a greater number of publications, mainly in journals, by researchers with lower PQ scholarship levels. These researchers have higher productivity in journals with higher Qualis indexes than researchers at higher PQ scholarship levels. To a lesser extent, this behavior is also noticed in conferences.

The differences between the average and median values demonstrate an imbalance in the number of papers published by each PQ scholarship level within each Qualis level. We can better see this balance when looking individually at the number of papers published by each researcher, as shown in Figure 8. We notice a great number of fellows who have a total number of publications higher than fellow researchers better ranked in the PQ scholarship than them. This behavior is also presented when considering higher strata in the Qualis classification. As an example, when we consider the four higher Qualis levels (A4, A3, A2, and A1), 45% of the PQ-2 researchers have a higher number of papers published than the average number of publications of the researchers in the PQ-1A level.

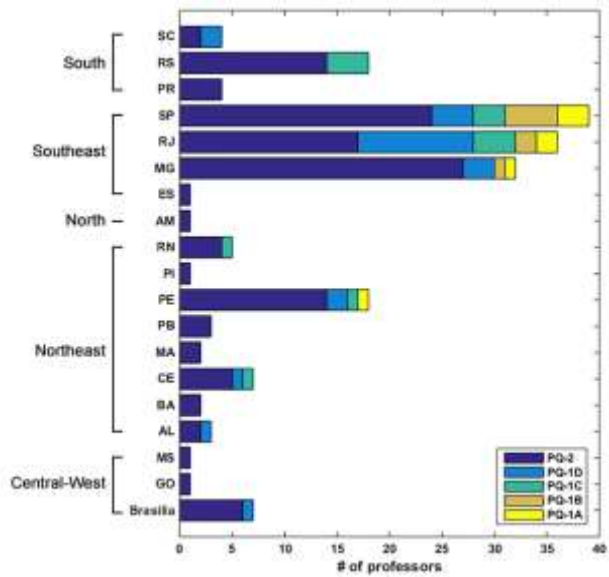

**Figure 4** – Distribution of PQ scholarships granted by state/region.

Figure 9 shows the distribution of the PQ scholarship holders' production in the four highest strata of Qualis (A4, A3, A2, and A1) concerning the year they obtained their Ph.D.. Researchers at lower PQ scholarship levels have shown a great effort to publish their work in more reputable journals, often supplanting researchers at higher productivity levels.

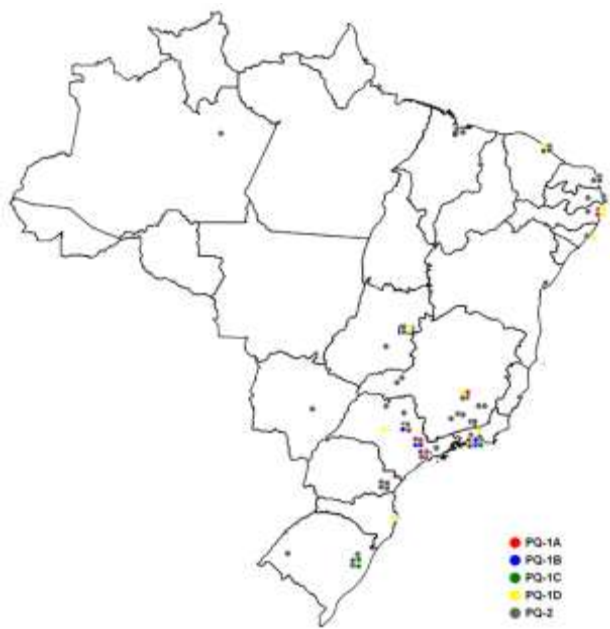

**Figure 5** – Map of the PQ scholarships granted.

Compared to journals, conferences are a more agile publishing venue. Most conferences have one single review step to determine whether a work should be accepted for publication or not, besides predetermined dates for submission and notification of acceptance. Differently, a journal may require multiple revisions over the time of one or more years. Therefore, there is a slightly superior number of papers published in conferences than in journals, as we can see in Figure 10. Over the 5 years (2017-2021), on average, for each paper published in a journal, a total of 1.57 papers are published in conferences. However, when we consider the four higher Qualis levels (A4, A3, A2, and A1), we notice that this difference decreases from 1 journal to 1.22 conferences.

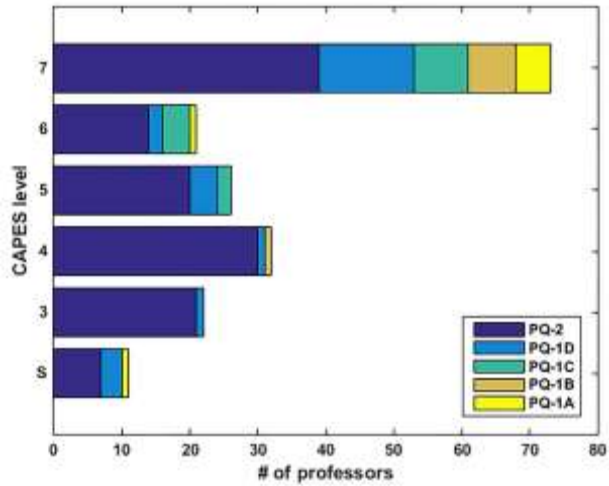

**Figure 6** – Distribution of PQ scholarships granted by CAPES level.

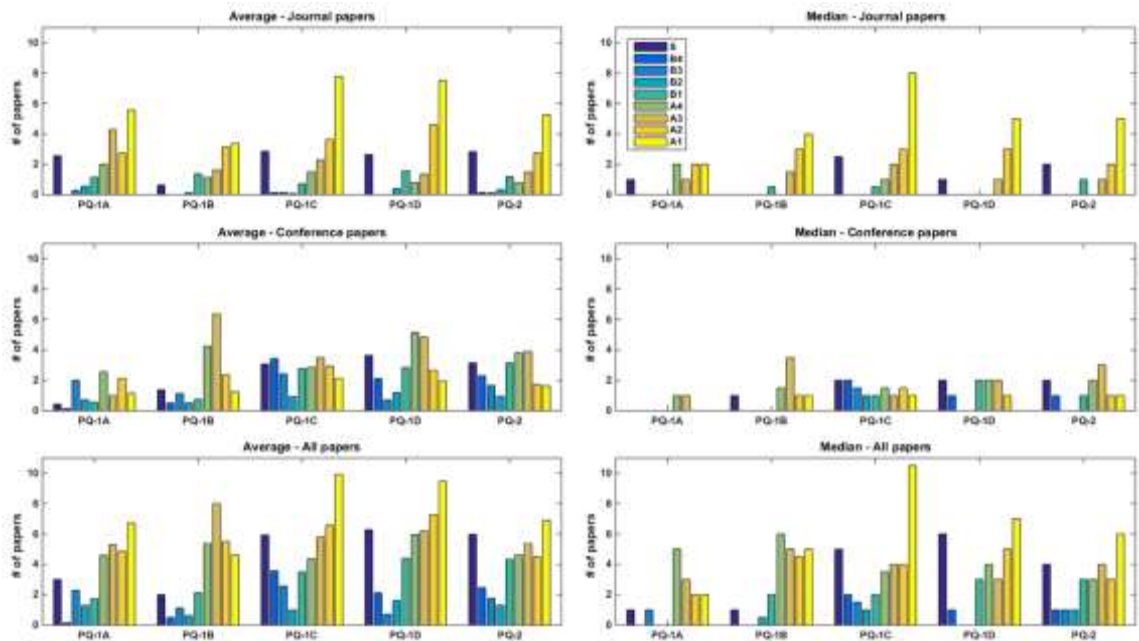

**Figure 7** – Average and median publications grouped by Qualis index and PQ scholarship rank (2017-2021)

We also evaluated the role of researchers in the formation of human resources, whether mentoring postgraduate students or undergraduate students (as in undergraduate research and term papers, presented at the end of the graduation course).

Differently from the analysis of publications, it can be seen that the number of students of each researcher has small variation for each PQ scholarship level, as shown in Figure 11. The most notable exceptions occur in the highest PQ scholarship level, PQ-1A, where one researcher presents an excessive number of students, while another researcher has none. In the lowest PQ scholarship level, PQ-2, we notice that the total number of students per researcher is unbalanced. Such a result is expected given the larger number of researchers in this group. Additionally, many of them have recently obtained their Ph.D. degrees. Thus, many did not have enough time as hired professors in a university to enter the graduate program or to mentor graduate students.

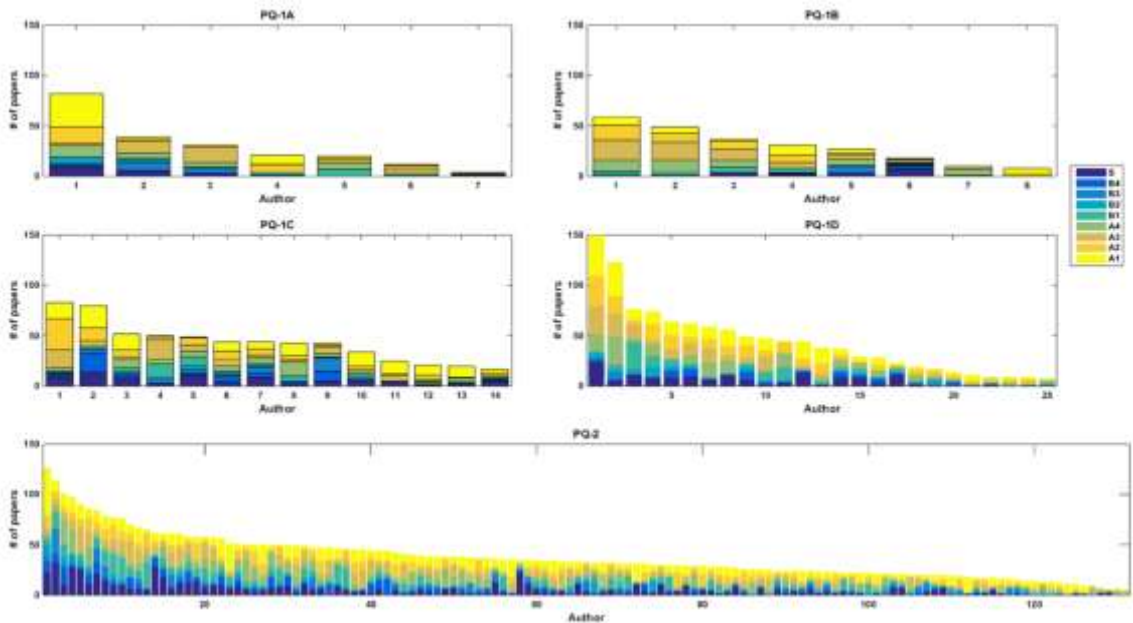

**Figure 8** – Number of paper of each researcher, ordered by quantity and grouped by Qualis index and PQ scholarship rank.

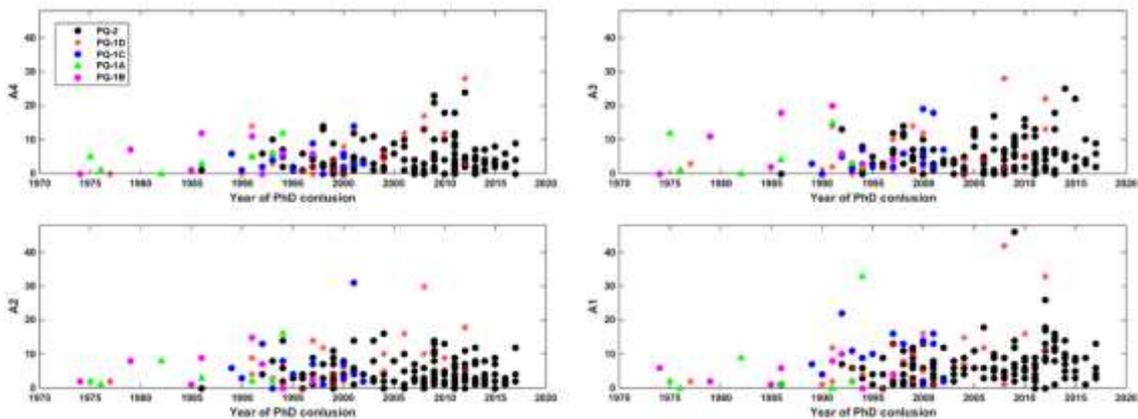

**Figure 9** – Scatterplot of the researchers' production according to the Qualis index and year of their Ph.D..

According to Figure 12, researchers at the highest PQ scholarship level dedicate themselves more to mentor graduate students. Even among them, it is possible to notice a small imbalance, mainly in the issue of ongoing mentoring. However, this result may just be due to problems with filling Lattes Résumés, as many researchers may choose to add this data only after completing the mentoring. Another important aspect is related to the mentoring of undergraduate students, which is mainly carried out by professors on the PQ-2 scholarship level.

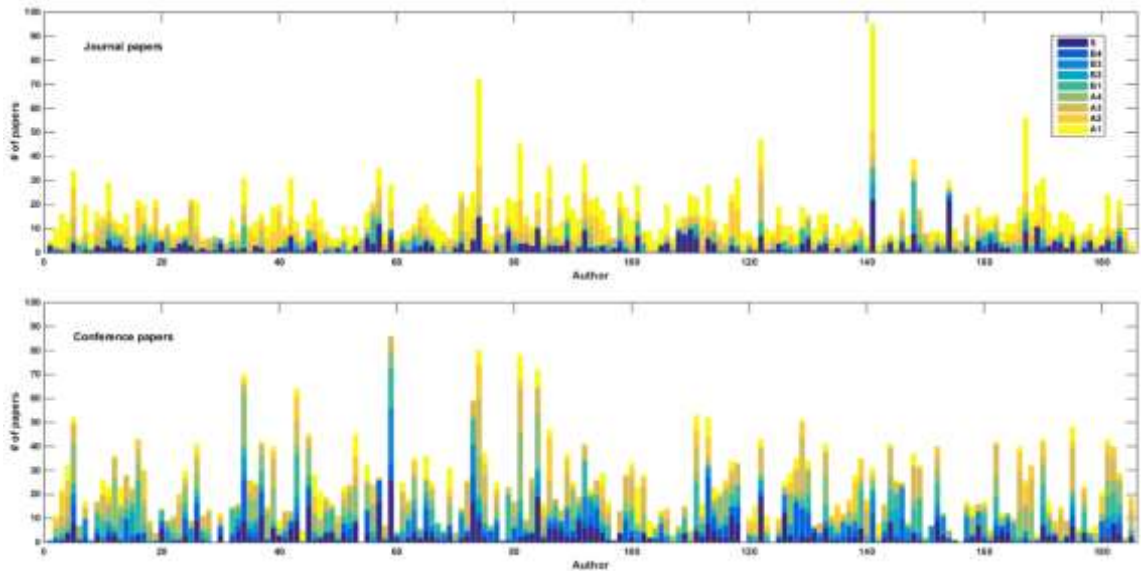

**Figure 10** – Number of papers published by researchers in each Qualis level from 2017 to 2021.

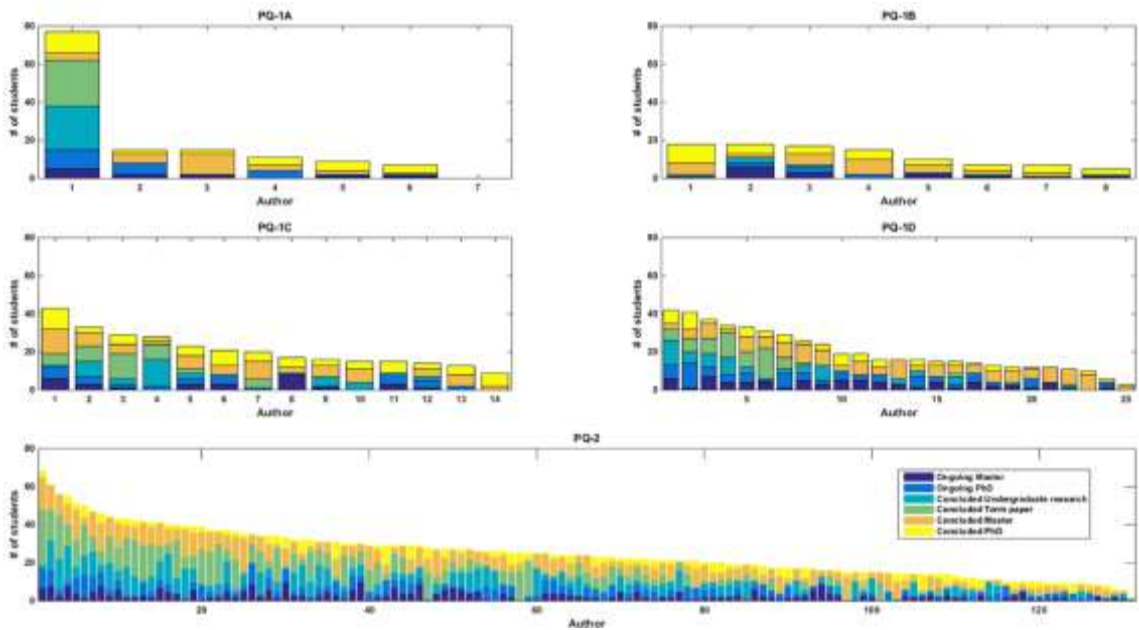

**Figure 11** – Number of students of each researcher, ordered by quantity and grouped by type of student and PQ scholarship rank.

Figure 13 shows the distribution of the number of students supervised by each researcher concerning the year they obtained their Ph.D. As previously discussed, most researchers at lower levels of the PQ fellowship have been actively engaged in mentoring undergraduate students, in the hopes that they become master students soon. PQ-2 researchers present a lower performance in the number of concluded Ph.D. This is expected due to the short career time that many have. In addition, many universities have rules for entry into graduate programs, as well as a minimum number of completed master's degree orientations so that they can mentor their first Ph.D. candidate.

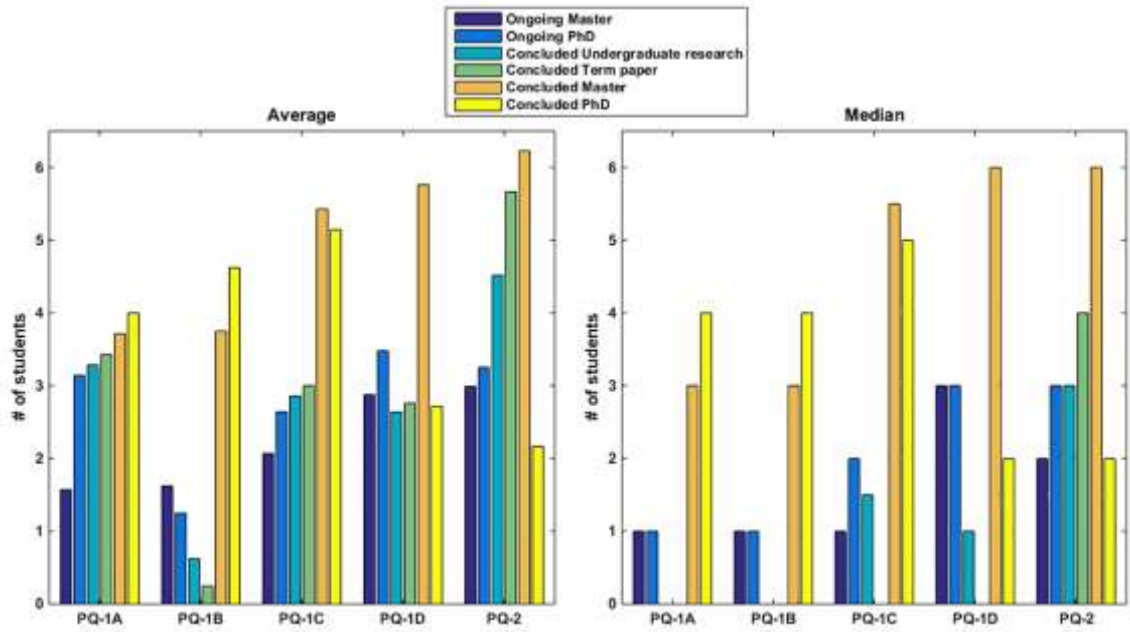

**Figure 12** – Average and median number of students grouped by type of student and PQ scholarship rank (2017-2021).

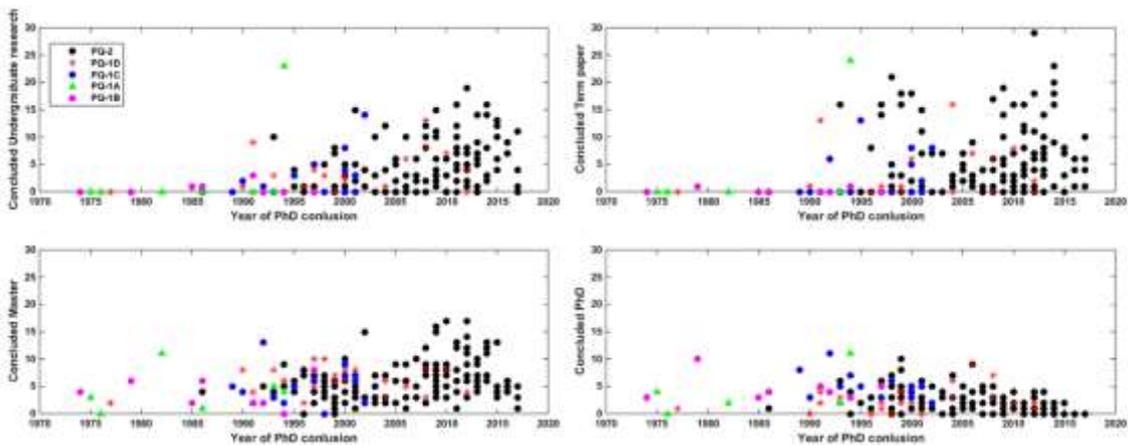

**Figure 13** – Scatterplot of the researchers mentoring activities according to the year of their Ph.D..

Finally, we analyzed the co-authorship network. To accomplish this task we extracted the titles of the papers in the Lattes Résumés of the selected researchers. The title of a research paper may contain errors (written with different characters, without accents, etc.), so we used Levenshtein distance to detect unique titles. From our total of 7,027 papers, we found 6,315 unique papers in 5 years. From this data, we were able to build a co-author network, where two researchers (nodes) are connected if they share a publication in common (edge).

Figure 14 shows the co-author network obtained. Most connected researchers are located at the same university or universities within the same region (e.g., São Carlos with UFSCar and USP/SC, Rio de Janeiro with UFF, UERJ, and LNCC). Regardless of the level of the PQ scholarship, there is a very large number of researchers without connections, an indication that there is little collaboration between these researchers. Another explanation for the low level of connections is due to the small number of Lattes Résumés (185) evaluated and the existence of many computer areas represented in these Résumés. Traditionally, areas such as software engineering and computer vision have a very low level of collaboration due to the distinction between the researched topics.

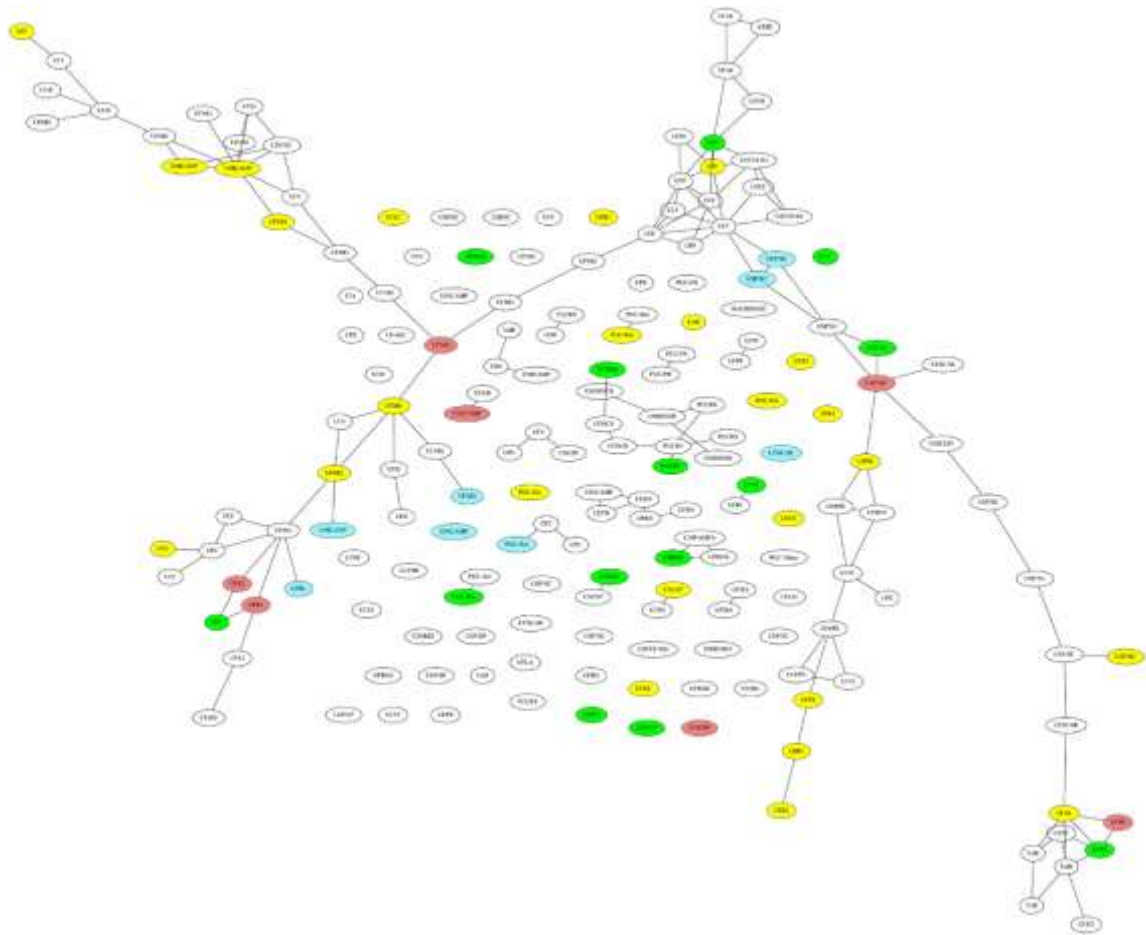

**Figure 14** – Co-author network computed for a 5-year period (2017-2021). PQ-1A: red; PQ-1B: blue; PQ-1C: green; PQ-1D: yellow; PQ-2: white.

## CONCLUSIONS

In this paper, we collected and evaluated the Lattes Résumés of the 185 researchers in the Computer Science area granted with PQ scholarship in the last notice. We evaluated the productivity of each professor, in each scholarship level, in terms of both quantity and quality. We used Qualis index, a classification of the journals and conferences according to its relevance, to measure the quality of the published research item. We found that higher PQ levels are associated with older faculty members, usually belonging to a level 7 CAPES postgraduate program, and located in the southeast region. This region is characterized by a higher level of development and IDH and it concentrates the majority of the scholarships (57.84%). We notice that researchers at lower PQ scholarship levels have shown a great effort to publish their work in more reputable journals, often supplanting researchers at higher productivity levels. In the formation of human resources, the number of students of each researcher does not vary so much in each PQ scholarship levels. Nevertheless, researchers at the highest PQ scholarship level dedicate themselves more to mentor graduate students.

## ACKNOWLEDGMENTS

André R. Backes gratefully acknowledges the financial support of CNPq (National Council for Scientific and Technological Development, Brazil) (Grant #307100/2021-9). Marcelo Keese Albertini gratefully acknowledges the financial support of CNPq (National Council for Scientific and Technological Development, Brazil) (Grant #406418/2021-7 and #306795/2022-1). This study was financed in part by the Coordenação de Aperfeiçoamento de Pessoal de Nível Superior - Brazil (CAPES) - Finance Code 001.

## AUTHOR CONTRIBUTIONS

ARB collected the data and created the figures. ARB and MKA performed data analysis and wrote the paper. All authors contributed toward interpreting the results, revising, and improving the paper.